\documentclass{elsart}

\usepackage{natbib}

\usepackage{graphicx}

\usepackage{amssymb}

\begin{document}

\begin{frontmatter}



\title{STM observation of initial growth of Sn atoms on Ge(001) surface}


\author{Kota Tomatsu\corauthref{cor1}}\ead{tomatsu@issp.u-tokyo.ac.jp}, \author{Kan Nakatsuji}, \author{Takushi Iimori}, \author{Fumio Komori}

\corauth[cor1]{Corresponding author. FAX: +81 4 7136 3474}
\address{Institute for Solid State Physics, University of Tokyo, Kashiwanoha, Kashiwa-shi, Chiba 277-8581, Japan}

\begin{abstract}
We have studied initial growth of Sn atoms on Ge(001) surfaces at room temperature and 80 K by scanning tunneling microscopy.
For Sn deposition onto the Ge(001) substrate at room temperature, 
the Sn atoms form two kinds of one-dimensional structures 
composed of ad-dimers with different alignment, in the $\langle 310 \rangle$ and the $\langle 110 \rangle$ directions, and epitaxial structures.
For Sn deposition onto the substrate at 80 K, 
the population of the dimer chains aligning in the $\langle 310 \rangle$ direction increases.
The diffusion barrier of the Sn adatom on the substrate kinetically determines
the population of the dimer chain. 
We propose that the diffusion barrier height depends on surface strain induced by the adatom.
The two kinds of dimer chains appearing on the Ge(001) and Si(001) surfaces with adatoms of the group-IV elements
are systematically interpreted in terms of the surface stain. 
\end{abstract}

\end{frontmatter}


\section{Introduction}\label{sec:intro}
It is known that group-IV (Si, Ge, Sn, and Pb) and group-III elements (Al, Ga, and In) on Ge(001) and Si(001) surfaces form one-dimensional (1-D) structures composed of adatom-dimers (dimer chains)
aligning in the $\langle 310 \rangle$ or $\langle 110 \rangle$ directions 
at their initial growth stage at room temperature (RT) \cite{galea2000,wulfhekel1997,qin1997,wingerden1997,veuillen1996,glueckstein1998,yang1995,nogami1991,baski1990,baski1991,falkenberg1997}.
Their length sometimes exceeds tens of nanometers while its width is a single-atom size. 
Thus, the dimer chain is a prototype of self-organization at surfaces, and 
its formation process and electronic states 
have been extensively studied over the past decade
for both scientific and technological interests. 
\par

On the clean Ge(001) and Si(001) surfaces,
the neighboring two surface atoms form a dimer to reduce the number of dangling bonds (DBs)
which increase surface energy \cite{koma1994}.
The dimer further lowers its energy by tilting (buckling) its dimer bond from the surface plane.
These dimers are formed into rows (dimer rows).
The buckling orientation of the dimer is
alternate in the direction perpendicular to the dimer axis, the dimer-row direction.
Similarly to the dimers on the clean surface,
the adatoms of group-IV elements on these surfaces form dimers for reducing the number of DBs,
and they are often buckled.
\par

Among the group-IV elements on the Ge(001) surfaces, however, 
little is known about initial growth of Sn.
Previous studies on the Sn/Ge(001) surfaces mainly focused on properties of Sn films
grown by molecular beam epitaxy, 
using Auger electron spectroscopy, Rutherford back scattering \cite{gossmann1987},
and atomic force microscopy \cite{dondl1997}. 
The structure of submonolayer Sn atoms, 
such as whether Sn atoms also form dimer chains or not, has not been reported so far.
\par

In the present paper, we demonstrate by means of scanning tunneling microscopy (STM) that
the Sn atoms form two kinds of dimer chains
aligning in the $\langle 110 \rangle$ and the $\langle 310 \rangle$ directions
similarly to Si/Si(001) and Ge/Si(001) surfaces \cite{qin1997,wingerden1997},
after Sn deposition onto the Ge(001) surface at RT and 80 K.
The ad-dimers in the dimer chain aligning in $\langle 310 \rangle$ direction
are intrinsically symmetric (or slightly tilted),
whereas those in the dimer chain aligning in the $\langle 110 \rangle$ direction
asymmetric.
At 80K, the number density of the dimer chains aligning in the $\langle 310 \rangle$ direction increases  while that in the the $\langle 110 \rangle$ direction decreases.
This suggests that the formation of the dimer chain is
governed by the diffusion barrier of the deposited adatom on the substrate as in the model proposed by Qin \textit{et al.} 
for Si/Si(001) and Ge/Si(001) surfaces \cite{qin1997}. 
We suggest that the diffusion barrier height depends on surface strain, 
and the existence of the two kinds of dimer chains on the Sn/Ge(001) surface is
interpreted in terms of strain induced by lattice mismatch between adatoms and substrate atoms.
\par

This paper is structured as follows.
In the next section, we describe the methods of our experiment.
We present the experimental results and discuss them in Section \ref{sec:results}.
The results for Sn deposition at RT and 80 K are shown
in Section \ref{sec:rtdepo} and \ref{sec:ltdepo}, respectively.
The formation process of the dimer chain is discussed in Section \ref{sec:formation}.
Our results are compared with those of the other group-IV elements in Section \ref{sec:comp}.
Finally, in Section \ref{sec:summary}, we summarize these results.

\section{Experiments}\label{sec:experiment}
We used two independent STMs for the observation at RT and at 80 K.
We acquired all the images in the present paper
in a constant-current mode with an electrochemically etched tungsten tip,
and analyzed with a homemade program.
\par

The substrate was cut from an n-type Ge(001) wafer (Sb-doped, 0.2-0.4 $\Omega $ cm)
and rinsed in ethanol and acetone. 
Then we introduced the substrate into an ultra-high vacuum (UHV) chamber
where the base pressure was maintained below $1 \times 10^{-8}$\,Pa.
A clean Ge(001) substrate was acquired by repeated cycles of Ar$^{+}$ sputtering for 20 min at 1 keV 
and d.c. annealing for 20 min at 920-1000 K in the UHV chamber.
Less than 0.1 monolayer (ML) of Sn was deposited onto the clean Ge(001) substrate at RT and 80 K
from an alumina crucible heated with a tantalum filament at a rate of 0.018 ML/min.
Here, ML is defined as the number density of Ge atoms at the clean Ge(001) surface,
and 1 ML = $6.25 \times 10^{18}/ \textrm{cm}^{2}$.
We monitored the deposition rate with a quartz crystal oscillator,
and calibrated it by counting the Sn atoms directly by STM.

\section{Results and discussion}\label{sec:results}
\subsection{Sn deposition at room temperature}\label{sec:rtdepo}
\subsubsection{Sn structures}\label{sec:surfstructure}

\begin{figure}[htbp]
    \begin{center}
        \includegraphics[keepaspectratio=true,scale=0.6]{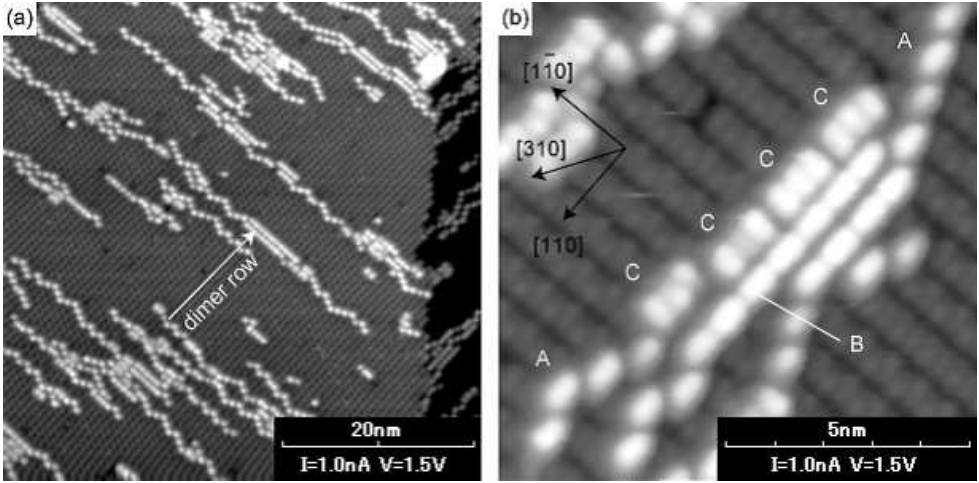}
    \end{center}
    \caption{(a) Empty-state image (tunneling current $I = 1.0$ nA and bias voltage $V = 1.5$ V) 
    of the Ge(001) surface covered with 0.07 ML of Sn deposited at RT.
    (b) Magnified image of the same surface in (a). 
    The Sn atoms form one-dimensional structures extending
    perpendicularly to the Ge dimer row.
    These surface structures are classified into three components (A, B, and C in (b)).
    The positions of the protrusions in the three components are schematically illustrated in Fig. \ref{fig:propos}.
    }
    \label{fig:rtsurf}
\end{figure}

Figure \ref{fig:rtsurf}(a) shows an empty-state image
of the Ge(001) surface covered with 0.07 ML of Sn deposited at RT.
We performed the STM measurement at RT. 
The arrow in the figure indicates the direction of the Ge dimer rows on the substrate.
On the surface, we can see 1-D structures extending perpendicularly to the Ge dimer rows. 
These 1-D structures tend to bunch, and inhomogeniously distribute on the substrate. 
Figure \ref{fig:rtsurf}(b) is a magnified image of the same surface as that in Fig. \ref{fig:rtsurf}(a). 
The 1-D structures consist of three basic components indicated by A, B, and C in the figure.

\par

\begin{figure}[htbp]
    \begin{center}
        \includegraphics[keepaspectratio=true,scale=0.6]{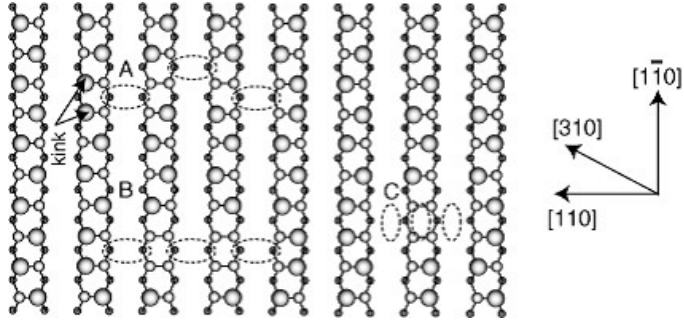}
    \end{center}
    \caption{Schematic illustration indicating the positions of the protrusions in the three components shown in Fig. \ref{fig:rtsurf}.
    The protrusions in the components A and B are located in the trough between the Ge dimer rows, 
    whereas those in the component C in the trough and on top of the Ge dimer rows.
    The two Ge dimers in the dimer rows adjacent to the component buckle in the same direction (kink).
    }
    \label{fig:propos}
\end{figure}

The positions of the protrusions in these three components with respect to the substrate 
are illustrated in Fig. \ref{fig:propos}.
The protrusions in the components A and B locate in the trough between the Ge dimer rows. 
In the component A, they align in the $\langle 310 \rangle$ direction,
that is, the separation between the two neighboring protrusions
in the dimer-row direction is $a$,
the lattice constant of the ($1 \times 1$) unit cell of the unreconstructed Ge(001) surface.
In the component B, on the other hand, they align perpendicularly to the Ge dimer rows, that is,
in the $\langle 110 \rangle$ direction. 
The component C is a cluster consisting of three protrusions: 
two of them are located in the trough and one on top of the Ge dimer row.  
All the protrusions in these three components have an oval-shape. 
Those in the components A and B are elongated perpendicularly to the dimer row,
whereas those in the component C in the dimer-row direction.
The results suggest that the atomic configuration of the component C
is different from those of the components A and B.
\par

The population densities of the three components
for the Ge(001) surface covered with 0.07 ML of Sn are
38 \%, 41 \% and 21 \% for the components A, B and C, respectively.
The component B appears mainly in the area where the three components bunch with one another.
The interval between the neighboring components B
at the bunch in the dimer-row direction is more than 2$a$.
On the other hand, the component C always appears next to
or inside the other components (see Fig. \ref{fig:rtsurf}(b)). 
The components A and B often coexist in a single isolated 1-D structure
having no adjacent 1-D structures in the dimer-row direction.
In such a case, long components A and B are formed.
\par

We note that the three components determine
the buckling orientation of the neighboring Ge dimers on the substrate.
The two Ge dimers in the dimer row adjacent to the component buckle in the same direction
as their atoms far from the component become upper: they are labeled with "kink" in Fig. \ref{fig:propos}.

\par

The empty-state image of the component A in Fig. \ref{fig:rtsurf}(b)
is similar to those of the dimer chains aligning in the $\langle 310 \rangle$ direction
reported for Ge/Ge(001) \cite{galea2000}, Si/Ge(001) \cite{wulfhekel1997},
Ge/Si(001) \cite{qin1997}, and Si/Si(001) \cite{wingerden1997} surfaces.
On the other hand, the empty-state image of the component B is
similar to those of the dimer chains aligning in the $\langle 110 \rangle$ direction
reported for Ge/Si(001) \cite{qin1997}, Si/Si(001) \cite{wingerden1997},
Pb/Si(001) \cite{veuillen1996}, Sn/Si(001) \cite{glueckstein1998},
and Pb/Ge(001) \cite{yang1995} surfaces.
These suggest that Sn atoms form a dimer chain on the Ge(001) substrate.

\subsubsection{Sn dimers}
We investigated the electronic states and the atomic configurations of the three components
by measuring their sample bias-voltage dependence of STM.
In the following, we will show the results for the component A, B, and C separately. 

\begin{figure}[htbp]
    \begin{center}
        \includegraphics[keepaspectratio=true,scale=0.6]{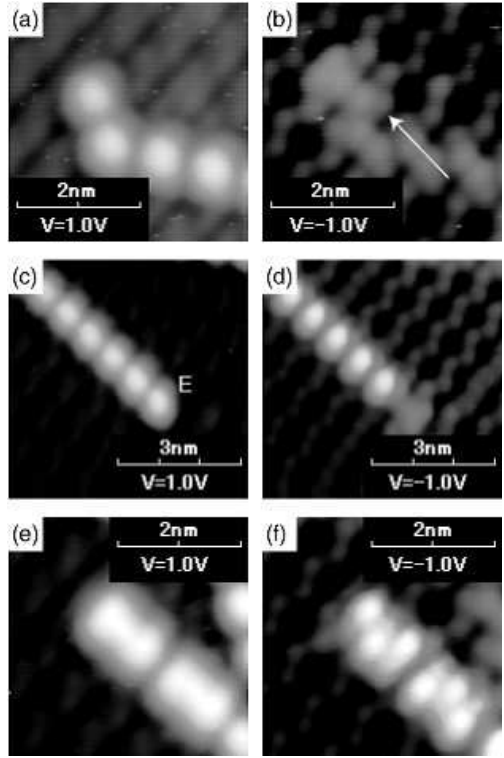}
    \end{center}
    \caption{Bias-voltage dependence of the three components: 
    (a) empty- and (b) filled-state images of the component A,
    (c) empty- and (d) filled-state images of the component B, and
    (e) empty- and (f) filled-state images of the component C.
    These images were recorded in a dual-bias mode at RT at $I = 1.0$ nA.  
    The line profile corresponding to the cross section indicated by the arrow in (b) 
    is shown in Fig. \ref{fig:linepro}.
    }
    \label{fig:bias}
\end{figure}

\begin{figure}[htbp]
    \begin{center}
        \includegraphics[keepaspectratio=true,scale=0.6]{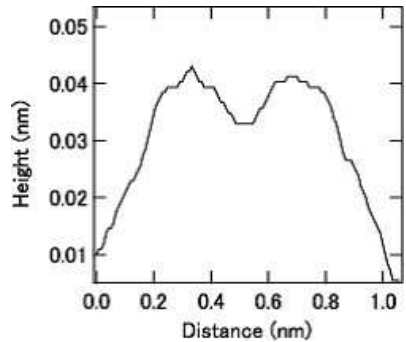}
    \end{center}
    \caption{Line profile of the protrusion in the component A 
    in the filled-state image shown in Fig. \ref{fig:bias}(b).
    The protrusion in the empty-state image splits into two protrusions.
    The Sn ad-dimer looks symmetric with respect to
    the center of the trough between the Ge dimer rows.
    }
    \label{fig:linepro}
\end{figure}

\paragraph*{Component A}\label{sec:310}
Figures \ref{fig:bias} (a) and (b) show the empty- and the filled-state images
of the same area including the component A.
In the empty-state image, the protrusions in the component A
are imaged $0.14$ nm higher than the lower atom of the substrate Ge dimer.
Here, we note that the lower atom of the Ge dimer is imaged
as a protrusion on the clean Ge surface in this condition.
Whereas, in the filled-state image, 
they are observed only $0.05$ nm higher than the upper atom of the substrate Ge dimer,
which is the protrusion on the clean Ge surface. 
Figure \ref{fig:linepro} shows line profile of this protrusion along 
the cross section indicated by the arrow in Fig. \ref{fig:bias}(b).
The single protrusion in the empty-state image splits
into two protrusions in the filled-state image. 
From these results, we can conclude that each protrusion in the component A 
is a Sn ad-dimer whose dimer axis is parallel to the substrate Ge dimer's.
STM studies on Ge/Ge(001), Si/Ge(001), Ge/Si(001), and Si/Si(001) surfaces
\cite{galea2000,wulfhekel1997,qin1997,wingerden1997} report
that the ad-dimer chains aligning in the $\langle 310 \rangle$ direction
were observed as very low protrusions in the filled-state image at RT as in the present case.

\par

\par

The line profile of the ad-dimer in Fig. \ref{fig:linepro} is symmetric
with respect to the center of the trough between the Ge dimer rows.
We also observed asymmetric protrusions in the component A. 
However their population was small, and less than 10 \% of all the protrusions in the component A
(their origin is discussed in Section \ref{sec:substitution}).
In general, there are two possible origins for such symmetric appearance of the dimer in STM images.
One is that its geometry is intrinsically symmetric,
and another is that an asymmetric dimer flip-flops back and forth thermally between
its two equivalent asymmetric geometries.
Such a flip-flopping dimer is imaged symmetric because of time average during the scanning of the surface typically in the order of a few tens msec/dimer.
The latter origin has been applied to the asymmetric Ge (Si) dimers on the clean Ge(001) (Si(001)) surface, which look symmetric in STM images at RT \cite{koma1994,kubby1987,sato1999}.
The observations at low temperature, where thermal flip-flop motion is inhibited, are essential for claiming the asymmetric dimer.
\par

We observed the component A at 80 K, and found that its ad-dimers are imaged symmetric.
Thus, we conclude that the ad-dimer in the component A is intrinsically symmetric,
or that even if it buckles, the tilting angle is quite small.
\par

Silva \textit{et al.} have calculated STM images for an almost symmetric Ge ad-dimer
in a trough between Si dimer rows on a Si(001) surface \cite{silva2001}.
In their calculations, similarly to our experimental results,
the Ge ad-dimer is imaged as single and double protrusions
in the empty- and the filled-state images, respectively.
Their shape and height also resemble those of the Sn ad-dimers in the component A.
In addition to the STM observation at 80 K,
this fact suggests that
the Sn ad-dimer in the component A is intrinsically symmetric (or slightly tilted).
\par

\begin{figure}[htbp]
    \begin{center}
        \includegraphics[keepaspectratio=true,scale=0.6]{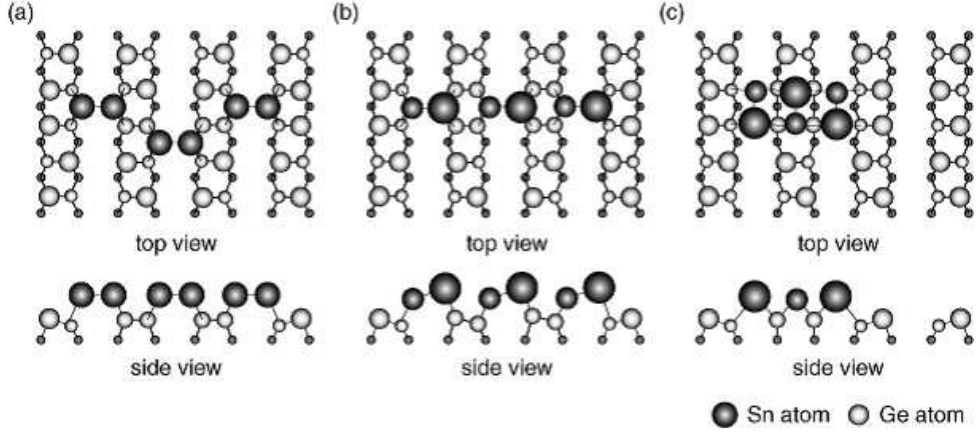}
    \end{center}
    \caption{Models of the atom configurations for (a) the component A, (b) B, and (c) C.
    The component A and B are composed of Sn ad-dimers whose dimer axes
    are parallel to the substrate Ge dimer's axis.
    On the other hand, the component C is composed of three Sn ad-dimers
    whose dimer axes are perpendicular to the substrate Ge dimer's axis.
    The Sn ad-dimers in the component A is intrinsically symmetric, or slightly buckled, 
    and those in the components B and C are intrinsically asymmetric.
    The component A and B are dimer chains aligning
    in the $\langle 310 \rangle$ and the $\langle 110 \rangle$ directions, respectively.}
    \label{fig:model}
\end{figure}

The model of the component A based on our STM images
is illustrated in Fig. \ref{fig:model}(a). 
The component A is composed of Sn dimers aligning in the $\langle 310 \rangle$ direction
whose dimer-axes are parallel to the substrate Ge dimer's axis.
This model is the same as that of the dimer chains extending in the $\langle 310 \rangle$ direction
reported previously for Ge/Ge(001), Si/Ge(001), Ge/Si(001), and Si/Si(001) surfaces
\cite{galea2000,wulfhekel1997,qin1997,wingerden1997}.
However, in these studies, the symmetry of the ad-dimer is not discussed so much.
In the present paper, we refer to the component A as a ``$\langle 310 \rangle$ dimer chain'' for convenience.

\paragraph*{Component B}\label{sec:110}
Figures \ref{fig:bias}(c) and (d) show the empty- and the filled-state images
of the component B in the same area recorded at RT. 
The protrusions in the component B are imaged
as high as a single step of the substrate in both empty- and filled-state images. 
In the filled-state image,
the $\langle 310 \rangle$ dimer chain is imaged as very low protrusions, as described above.
Hence, the electronic states of the component B are
quite different from those of the $\langle 310 \rangle$ dimer chain.
On the other hand, in the filled-state image, the protrusion at the edge of the component B 
(indicated by E in Fig. \ref{fig:bias}(c)) looks
as low as the ad-dimer in the $\langle 310 \rangle$ dimer chain, and
splits into two protrusions.
This result indicates that the protrusion at the edge is a Sn ad-dimer.
Whether this edge ad-dimer is symmetric or not is open to discussion
because it is not clearly imaged owing to
the adjacent kink in the substrate Ge dimer row and the adjacent protrusion in the component B.
\par

\begin{figure}[htbp]
    \begin{center}
        \includegraphics[keepaspectratio=true,scale=0.6]{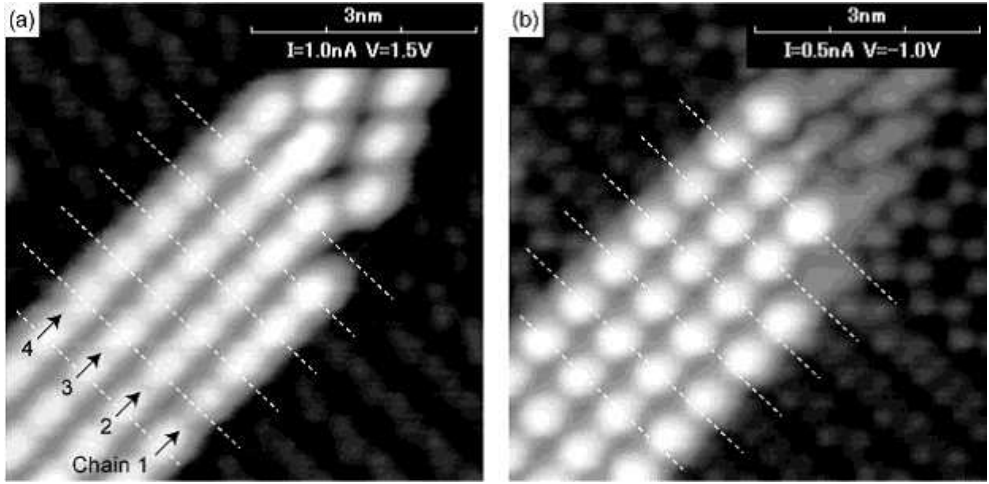}
    \end{center}
    \caption{Bias-voltage dependence of the component B: 
    (a) empty- and (b) filled-state images.
    The dotted lines in the figure indicate the 
    center of the trough between the Ge dimer rows underneath the dimer chains.
    In the filled-state image, the protrusions in Chain 1
    shift from their positions in the empty-state image
    in the direction perpendicularly to the substrate dimer rows.
    On the other hand, Chains 2-4 do not shift.}
    \label{fig:110bias}
\end{figure}

Figure \ref{fig:110bias} shows the bias-voltage dependence of the component B bunching with each other. 
We notice difference in the position of the protrusions between the filled- and the empty-state images. 
The dotted lines in the figure indicate the center of the trough
between the neighboring Ge dimer rows on the substrate.
In the empty-state image, all the protrusions are located at just the center of these lines, 
whereas in the filled-state image, the protrusions in Chain 1 in the figure
shift by $\sim 0.5 a$ perpendicularly to the dimer rows.
We can recognize this by comparing Chain 1 with Chain 2, which does not shift.
The direction of the shift tends to be the same among the protrusions in the same chain.
We found that about half of protrusions in the component B shift in the filled-state image at RT,
whereas in the observation at 80 K, almost all does.

\par

Previous STM studies on group-IV elements on the Ge(001) and Si(001) surfaces
have reported that parts of protrusions originating from the ad-dimers in the dimer chain
aligning in the $\langle 110 \rangle$ direction shift perpendicularly to the substrate dimer row
in the filled-state image although they are located
just at the center of the trough in the empty-state image \cite{glueckstein1998,yang1995}.
The observed shift is attributed to the existence of an asymmetric ad-dimer in the trough in the following way.
\par

In the asymmetric ad-dimer, whose axis is parallel to the substrate dimer's axis,
the electrons transfer from the $\pi$-state of the lower atom to that of the upper atom.
Consequently, in the filled-state image, the upper atom is imaged higher than the lower atom
because of two reasons;
the upper atom locates at geometrically higher position with respect to the surface than the lower atom,
and the local density of filled states is higher at the upper atom than at the lower atom.
On the other hand, in the empty-state image, the geometrical effect compensates the electronic effect
(i.e., higher local density of empty states at the lower atom than at the upper atom),
and image heights of the upper and lower atoms are almost equal to each other.
This can be applied to the present results of the bias voltage dependence shown in Fig. \ref{fig:110bias}.
Thus, we conclude that the component B is
a chain of asymmetric Sn ad-dimers aligning in the $\langle 110 \rangle$ direction. 
\par

Here, we note that this is not the case in the asymmetric dimers of the clean Ge(001) and Si(001) surfaces,
where the lower atom is imaged higher than the upper atom in the empty-state images.
For these surfaces, the protrusion originating from the dimer shifts
to the opposite direction to that in the filled-state images.
The apparent height difference between the two atoms of the dimer in the image
depends on the differences in both their real height and their local density of states.
The difference in the bias voltage dependence between the Sn ad-dimer
and the dimers of the clean Ge(001) and Si(001) surfaces
should be explained in terms of difference in balance between the electronic and the geometrical effects.
\par

Figure \ref{fig:model}(b) shows the model of the atomic configuration of the component B on the basis of our STM observations.
The component B consists of asymmetric Sn ad-dimers in the trough
whose dimer axes are parallel to the substrate Ge dimer axis.
The ad-dimers in the same dimer chain buckle in the same direction,
and parts of them flip-flop at RT.
In the filled-state image, these upper atoms are imaged higher,
and therefore the protrusion originating from the ad-dimer shifts from the center of the trough.
In the following, we refer to the component B as a `` $\langle 110 \rangle$ dimer chain'' 
to distinguish it from the $\langle 310 \rangle$ dimer chain.

\paragraph*{Component C}\label{sec:tri}
The empty- and filled-state images of the component C recorded at RT are
shown in Figs. \ref{fig:bias}(e) and (f), respectively.
In the empty-state image, the protrusions in the component C align in line, 
whereas, in the filled-state image, in zigzag. 
It seems that, similarly to the $\langle 110 \rangle$ dimer chain,
this zigzag appearance in the filled-state image results from asymmetric Sn ad-dimers.
\par

Figure \ref{fig:model}(c) illustrates our model of the atomic configuration of the component C. 
The component C is composed of Sn dimers whose axes are perpendicular
to the Ge dimer axis on the substrate. 
Because this Sn ad-dimer buckles oppositely to its neighboring ad-dimers, 
the filled states of the upper-dimer atoms are imaged as zigzag protrusions. 
On the other hand, in the empty-state image, the buckling of the ad-dimer is
smeared by the spatial effect mentioned in the previous subsection,
and the protrusions are imaged in line. 
Note that the component C is an epitaxial structure on the Ge substrate.
Its atomic configuration is the same as that of the Ge dimers 
on the upper terrace of the Ge(001) substrate.
We rarely observed epitaxial structures composed of more than three Sn ad-dimers.
This seems to be due to strain induced by a lattice mismatch between the Sn and Ge crystals.
\par

The population density of the component C is smaller than those of the $\langle 310 \rangle$
and the $\langle 110 \rangle$ dimer chains as mentioned in Section \ref{sec:surfstructure}. 
This suggests that the energy barrier for the formation of the component C is
larger than those of the other components.
For the formation, the Sn adsorbates have to cut off the Ge dimer bonds on the substrate
as shown in the side-view image in Fig. \ref{fig:model}(c).
The large energy barrier for the formation of the component C
seems to originate from the large energy for cutting off the Ge dimer bond.
Hereafter, we call the component C a `` triplet dimer''.

\subsubsection{Substitution between Sn and Ge atoms}\label{sec:substitution}
\begin{figure}[htbp]
    \begin{center}
        \includegraphics[keepaspectratio=true,scale=0.6]{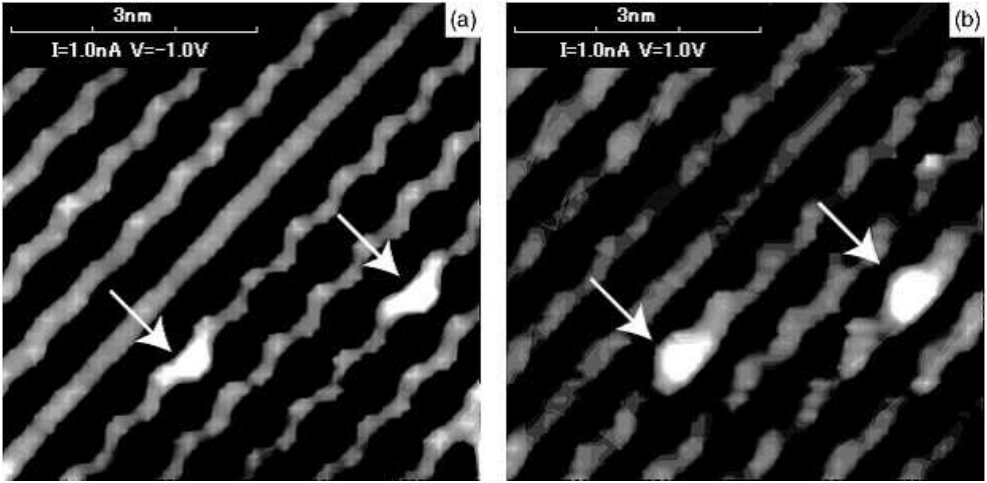}
    \end{center}
    \caption{(a)Filled- and (b) empty-state images of the heterogeneous dimer at the substrate Ge atom
    position at room temperature.
    We acquired the images in a dual-bias mode. 
    The arrows indicate the positions of the heterogeneous dimers.
    The empty states (filled states) of the heterogeneous dimer are imaged higher than those of the substrate Ge dimer are.
    }
    \label{fig:innerpro}
\end{figure}

When we deposit the Sn atoms onto the Ge substrate at RT,
we notice that, at the substrate Ge dimer positions,
there are dimers whose electronic states are different from those of the Ge dimer.
Figures \ref{fig:innerpro} (a) and (b) show filled- and empty-state images of such a heterogeneous dimer.
We indicate their positions by the arrows in the figures.
The upper (lower) atom of the heterogeneous dimer is
0.03 nm (0.02 nm) higher than that of the substrate Ge dimer in the filled-state (empty-state) image.
The flip-flop motion of the substrate Ge dimer is inhibited in the dimer row containing the heterogeneous dimer.
For 0.07 ML of Sn deposition, the population of the heterogeneous dimers was less than 0.01 ML.
It further increases after annealing this surface at 420 K.
Hence, we can rule out possible impurities in the Ge wafer or the crucible.

\par

We propose that the heterogeneous dimer originates from a mixed dimer composed of Sn and Ge atoms (Sn-Ge dimer)
at the Ge dimer position on the substrate for the following reasons.
Firstly, group-IV elements sometimes exhibit intermixing
on the Ge(001) and Si(001) surfaces during deposition at RT.
For example, on Ge deposited Si(001) surfaces, the Ge atoms intermix with the substrate Si atoms,
and Ge-Si dimers are formed at the Si dimer position \cite{patthey1995,chen1997,gay1999,qin2000may,qin2000oct}.
As discussed later in Section \ref{sec:comp},
the lattice mismatch at a crystal interface is almost the same between for Sn/Ge(001) and Ge/Si(001) surfaces,
and both the $\langle 310 \rangle$ and the $\langle 110 \rangle$ dimer chains are
formed in these systems.
It is quite likely that the Sn-Ge dimer is formed in the Ge dimer rows as on the Ge/Si(001) surface.
Secondly, the formation of the mixed dimer can explain the result
that the heterogeneous dimers inhibit the flip-flop motion of the substrate Ge dimer row around them.
The Sn-Ge dimer is chemically asymmetric,
and its two kinds of buckled geometries are energetically inequivalent.
Therefore, it seems that either buckled geometry is considerably stable, 
and the Sn-Ge dimer does not flip-flop even at RT.
\par

In Section \ref{sec:surfstructure},
we pointed out that less than 10 \% of the protrusions
in the $\langle 310 \rangle$ dimer chains are asymmetric. 
This also supports the above proposal.
The substitution of Sn atoms in the Ge substrate suggests that the existence of 
Sn-Ge or Ge ad-dimers on the substrate.
The asymmetric ad-dimer in the $\langle 310 \rangle$ dimer chain
is considered to be an asymmetric Sn-Ge ad-dimer.
We infer that the $\langle 110 \rangle$ dimer chain and the triplet dimer also contain 
the Sn-Ge or Ge ad-dimers.

\subsection{Sn deposition at low temperature}\label{sec:ltdepo}
Deposited atoms diffuse on a surface just after deposition.
This kinetically determines a morphology of adatom islands \cite{qin1997,brocks1993}.
Thus, we can get information on the formation process and the stability of the dimer chains by changing the substrate temperature during the deposition.
Here we discuss results for Sn deposition onto the surface at low temperature.
\par

\begin{figure}[htbp]
    \begin{center}
        \includegraphics[keepaspectratio=true,scale=0.6]{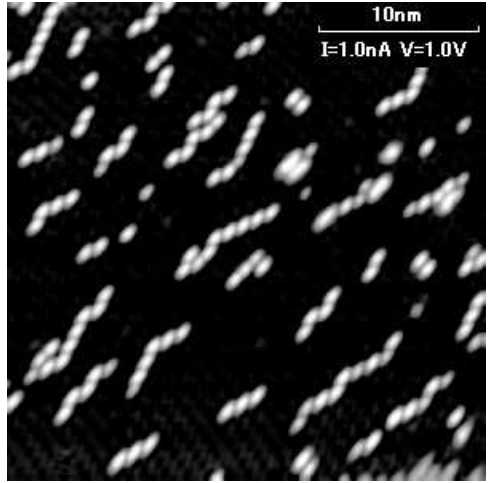}
    \end{center}
    \caption{Empty-state image of the Ge(001) surface after Sn deposition at 80 K.
    Sn coverage is 0.07 ML, and we recorded the image at RT.
    The dimer chains are shorter than those after the Sn deposition at RT.
    The population of the $\langle 310 \rangle$ dimer chains is larger than that
    after RT deposition.
    }
    \label{fig:ltdepo}
\end{figure}

Figure \ref{fig:ltdepo} shows the empty-state image of 
the Ge(001) surface at RT after Sn deposition at 80 K.
The three components (i.e., the $\langle 310 \rangle$ and the $\langle 110 \rangle$ dimer chains,
and the triplet dimer) are formed as for the RT deposition.
We notice increase of short dimer chains.
Even single Sn ad-dimers, we rarely observed for the Sn deposition at RT, are formed
on the surface in this case. 
\par

\begin{figure}[htbp]
    \begin{center}
        \includegraphics[keepaspectratio=true,scale=0.6]{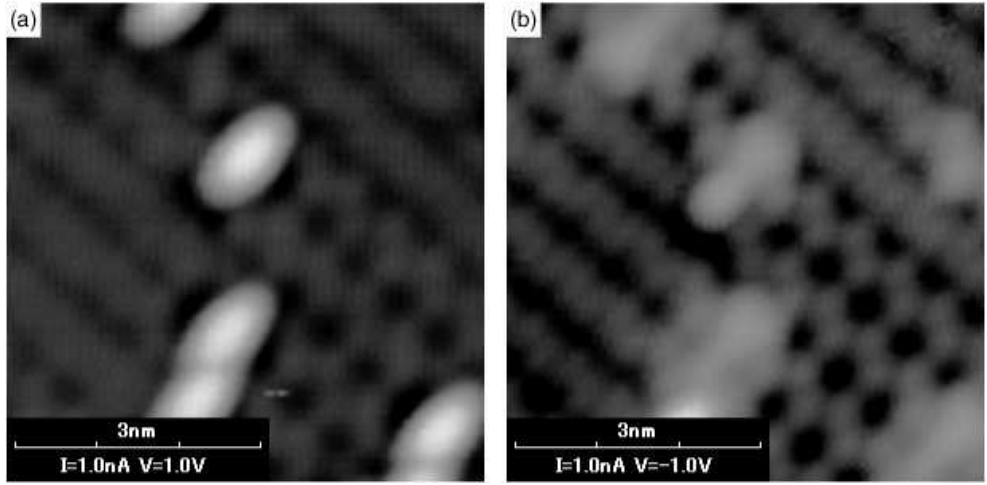}
    \end{center}
    \caption{(a)Empty- and (b)filled-state images of the same single ad-dimer at RT.
    In the empty-state image,
    the single dimer is imaged as a protrusion of monolayer height, 
    whereas in the filled-state image,
    it looks very low and asymmetric with respect to the trough of the Ge dimer rows.
    }
    \label{fig:single}
\end{figure}

The single ad-dimer is located in the trough, 
and its dimer axis is parallel to the substrate Ge dimer's axis.
Figure \ref{fig:single} shows the bias-voltage dependence of the single ad-dimer.
Similarly to the Sn ad-dimers in the $\langle 310 \rangle$ dimer chain,
the empty states of the single dimer are imaged as a protrusion of monolayer height,
and its filled states are imaged as a low protrusion.
The filled states look asymmetric with respect to the center of the trough.
A kink is formed in one of the two Ge dimer rows adjacent to the single ad-dimer.
This result differs from those in the previous studies on Ge/Ge(001) surfaces,
where the kink is formed in both of substrate dimer rows adjacent to the single Ge ad-dimer
\cite{zandvliet2000,galea2000sep}.
We suppose that the single Sn ad-dimer is trapped by a defect underneath it, such as a dimer vacancy,
and the single Sn ad-dimer is not stabilized on the ordered Ge substrate at RT.

\par

\par

The population densities of the three components are different between RT and 80 K depositions.
For the RT deposition, the population density of the $\langle 310 \rangle$ dimer chains
is almost the same as that of the $\langle 110 \rangle$ dimer chains
as described in Section \ref{sec:rtdepo}.
On the other hand, for the 80 K deposition, it is much larger.
For example, when we deposited 0.07 ML of Sn onto the substrate at 80 K,
the population densities of the $\langle 310 \rangle$ dimer chains, 
the $\langle 110 \rangle$ dimer chains, and the triplet dimers were
75 \%, 14 \%, and 11 \%, respectively.
These results indicate that the activation energy
for the formation of the $\langle 110 \rangle$ dimer chain is higher than
that of the $\langle 310 \rangle$ dimer chain.
We will discuss the difference in the activation energy between the two kinds of dimer chains
in the next section in conjunction with their formation process.

\subsection{Formation process of dimer chain}\label{sec:formation}
For the formation process of the Al dimer chains on the Si(001) surfaces, 
the surface polymerization mechanism was proposed by Brocks \textit{et al.} \cite{brocks1993}.
In their model, attractive interaction between the ad-dimers originates from
the presence of DBs of the substrate dimer.
On the Al/Si(001) surfaces, only the $\langle 110 \rangle$ dimer chain is formed at RT,
and the existence of the $\langle 310 \rangle$ dimer chain was not considered at all in their model.
On the other hand, on Si/Si(001) and Ge/Si(001) surfaces, 
both $\langle 110 \rangle$ and $\langle 310 \rangle$ dimer chains are formed near RT,
and the $\langle 110 \rangle$ dimer chain becomes predominant above RT ($\sim400$ K)
\cite{qin1997,wingerden1997}.
Hence, Qin \textit{et al.} have incorporated kinetics of adatoms
into the Brocks \textit{et al.}'s model to explain the existence of the $\langle 310 \rangle$ dimer chains near RT \cite{qin1997}.
\par

\begin{figure}[htbp]
    \begin{center}
        \includegraphics[keepaspectratio=true,scale=0.6]{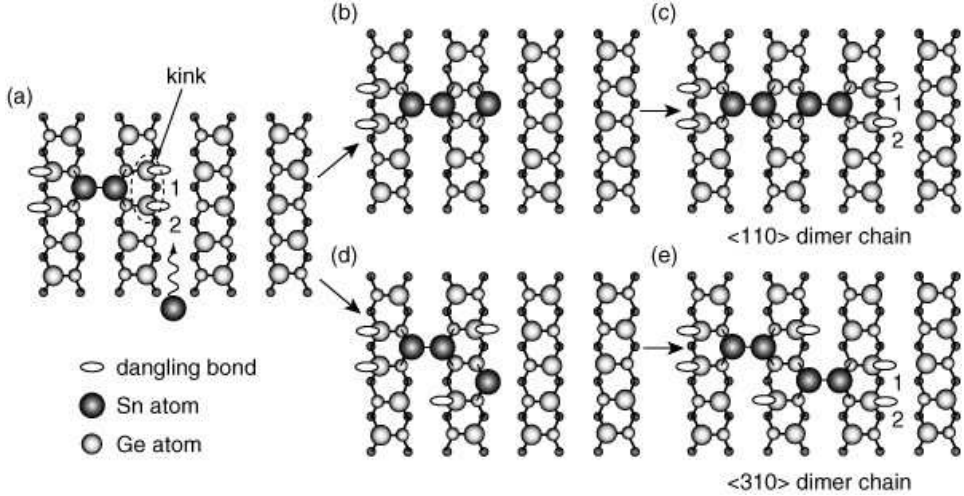}
    \end{center}
    \caption{Schematic illustrations of the formation processes of 
    the $\langle 110 \rangle$ dimer chain ((a)$\to$(b)$\to$(c)) 
    and the $\langle 310 \rangle$ dimer chain ((a)$\to$(d)$\to$(e)) on the basis of the Qin's model.
    For details, see the main text.}
    \label{fig:qin}
\end{figure}

Applying Qin \textit{et al.}'s model to the Sn/Ge(001) surface,
we can understand the predominance of the $\langle 310 \rangle$ dimer chain at 80 K
described in the previous section.
Figure \ref{fig:qin} illustrates the formation process of the Sn dimer chain on the Ge(001) surface on the basis of this model.
First, two Sn adatoms diffusing on the substrate form a single ad-dimer in the trough (Fig. \ref{fig:qin}(a)).
On the clean Ge(001) surfaces,
the two DBs of the Ge dimer form a $\pi$-bond to reduce the surface energy.
When the Sn dimer adsorbs in the trough, 
the $\pi$-bonds of the Ge dimers underneath it are cut off for formation of Sn-Ge bonds, 
and the created unpaired DBs give reactive sites for the diffusing Sn adatoms as schematically indicated by the ovals in the figure.
It is expected that, similarly to the other group-IV elements
\cite{zandvliet2000,galea2000sep,swartzentruber1996,yamasaki1996,borovsky1999,lu2000},
the Sn adatoms diffuse along the substrate Ge dimer rows. 
Then, the third Sn adatom diffusing along the Ge dimer row adsorbs
at the site 1 or the site 2 (Fig. \ref{fig:qin}(b) or (d)), and form an ad-dimer with the fourth Sn adatom.
Thus, the dimer chain grows in the $\langle 110 \rangle$ or $\langle 310 \rangle$ directions (Fig. \ref{fig:qin}(c) or (e)). As shown in Section \ref{sec:rtdepo},
the Sn ad-dimer fixes the buckling orientation of the substrate Ge dimers adjacent to it,
and the kink is formed in the dimer row. The kink makes the diffusion barrier from the site 2 to 1 high \cite{qin1997,liu2000}. At RT, the Sn atom arriving at the substrate by the deposition can diffuse from the site 2 to 1 before loosing its kinetic energy. On the other hand, at 80 K, the Sn adatom rapidly looses its kinetic energy,
and few of them can move to the site 1. This prevents the formation of the $\langle 110 \rangle$ chain at low temperature. 
\par

The adsorption energy at the site 1 is higher than that at the site 2 because the number of the DBs on the substrate
under the $\langle 310 \rangle$ dimer chain is
larger than that under the $\langle 110 \rangle$ dimer chain:
the four substrate Ge dimers adjacent to the Sn ad-dimer
in the $\langle 110 \rangle$ and the $\langle 310 \rangle$ dimer chains
have zero and two DBs, respectively.
Hence, the population of the $\langle 110 \rangle$ dimer chains increases at RT
(as described in Section \ref{sec:rtdepo}).

\par

However, it is noted that the Qin \textit{et al.}'s model should be modified
to explain the observed long dimer chains at RT. 
In Section \ref{sec:rtdepo}, we showed that
the $\langle 110 \rangle$ and the $\langle 310 \rangle$ parts in the same dimer chain
tends to grow continuously, and the long $\langle 110 \rangle$
and the long $\langle 310 \rangle$ dimer chains coexist at RT. 
In other word, once the dimer aligns in the $\langle 110 \rangle$ direction,
successive incorporations of the ad-dimer in the $\langle 110 \rangle$ direction take place.
The successive incorporation in the $\langle 110 \rangle$ direction 
is inconsistent with the Qin \textit{et al.}'s model, 
where there is no difference in the diffusion barrier height from the site 2 to 1 between the $\langle 110 \rangle$ and the $\langle 310 \rangle$ dimer chains.
Thus, in their model, the $\langle 110 \rangle$ and the $\langle 310 \rangle$ parts
should appear in the same dimer chain
at a constant probability depending on temperature.
Here, we propose that this difference in the diffusion barrier height
originates from the difference in structural symmetricity of the Sn ad-dimer.
\par

\begin{figure}[htbp]
    \begin{center}
        \includegraphics[keepaspectratio=true,scale=0.6]{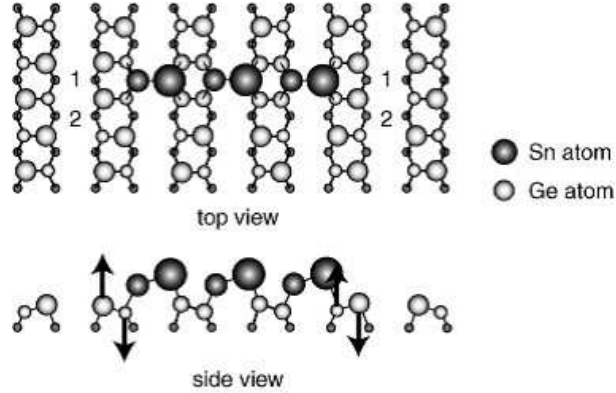}
    \end{center}
    \caption{Schematic illustration showing the mechanism for the reduction of the diffusion barrier height, which is caused by the ad-dimer buckling for the $\langle 110 \rangle$ dimer chain.
    When the Sn ad-dimers buckle, the kinks adjacent to the dimer chain 
    shift perpendicularly to the surface in the direction indicated by the arrows. 
    This decreases the diffusion barrier height from the site 2 to 1.}
    \label{fig:buckled110}
\end{figure}

In Section \ref{sec:rtdepo}, we showed that the $\langle 310 \rangle$ dimer chain consists of
intrinsically symmetric ad-dimers (or slightly buckled ad-dimers),
whereas the $\langle 110 \rangle$ dimer chain asymmetric ad-dimers.
Buckling orientation of the ad-dimers in the $\langle 110 \rangle$ dimer chain
tends to be the same to minimize the lattice strain energy.
From simple consideration of the surface lattice strain,
the edge ad-dimer and the adjacent inner ad-dimer
should buckle in the same direction
through the oppositely buckled substrate dimer between them
as shown in Fig. \ref{fig:model}(b) (see also Fig. \ref{fig:buckled110}) to relax surface strain.
Unfortunately, this is experimentally not comfirmed because the filled states of
the edge ad-dimer of the $\langle 110 \rangle$ dimer chain are not so clearly imaged
in contrast to the other inner ad-dimers in the same dimer chain.

\par

The difference in the ad-dimer buckling
between the $\langle 110 \rangle$ and the $\langle 310 \rangle$ dimer chains
accounts for the difference in the diffusion barrier height as follows.
When the Sn ad-dimer in the $\langle 110 \rangle$ dimer chain buckles
as the outside atom of the dimer chain becomes the lower (upper) atom, 
the upper atoms of the Ge kink adjacent to the dimer chains is shifted upward (downward) 
(see Fig. \ref{fig:buckled110}).
Then, the kink shift changes the diffusion barrier height of the Sn adatom.
The barrier height should decrease at either side of the dimer chain in the figure.
The buckling orientation of the newly adsorbed Sn ad-dimer is
the same as that of the other ones.
Hence, the diffusion barrier height always becomes lower, 
and successive incorporations in the $\langle 110 \rangle$ direction take place.
\par

Here, one might wonder about the result that the inner
and the edge ad-dimers in the $\langle 110 \rangle$ dimer chain
have different electronic states despite of their geometric similarity.
On the other hand, one might wonder that the edge ad-dimer in the $\langle 110 \rangle$ dimer chain
and the ad-dimer in the $\langle 310 \rangle$ dimer chain
have similar electronic states despite of their geometrical difference.
There are four substrate Ge dimers adjacent to the ad-dimer in the dimer chain.
The four Ge dimers adjacent to the edge ad-dimer in the $\langle 110 \rangle$ dimer chain
and those adjacent to the ad-dimer in $\langle 310 \rangle$ dimer chain have two DBs.
On the other hand, those adjacent to the inner ad-dimer in the $\langle 110 \rangle$ dimer chain
have no DB (see Fig. \ref{fig:qin}).
We suppose that the difference in the electronic states of the ad-dimers
originates not from their symmetricity, but from the presence of the DBs in the substrate Ge dimers under them.
\par

Tight-binding and \textit{ab-initio} calculations on the Ge/Si(001) surface have reported that,
as in the case of the Sn/Ge(001) surface,
filled states of the Ge ad-dimers in the $\langle 310 \rangle$ dimer chain or those of a single Ge ad-dimer
are weakly imaged \cite{silva2001, liu2000}.
Silva \textit{et al.} have attributed such lower appearance of the Ge ad-dimer in the filled-state image
to localization and absence of filled states near the Fermi energy
at the DBs of the substrate Si dimer and at the Ge ad-dimer, respectively \cite{silva2001}.
Because of structural similarity between the Sn/Ge(001) and the Ge/Si(001) surfaces (see Section \ref{sec:comp}),
the same seems to be applied to the present case.
We expect that the presence of DBs in the substrate Ge dimer
accounts for the lower density of filled states of the Sn ad-dimer in the $\langle 310 \rangle$ dimer chain. 

\subsection{Comparison with the other group-IV elements on Si(001) and Ge(001) surfaces}\label{sec:comp}
We have demonstrated that Sn atoms 
form dimer chains on Ge(001) surfaces after Sn deposition at RT similarly to the other group-IV elements.
Here we will compare the Sn/Ge(001) surface with the other surfaces.
\par

As mentioned in the introduction,
when group-IV elements are deposited onto the Si(001) or Ge(001) surfaces at RT,
the $\langle 110 \rangle$ dimer chain or the $\langle 310 \rangle$ dimer chain are formed
depending on the combination of the elements.
We can classify these surfaces into three groups by the kind of dimer chain:
a surface with only the $\langle 310 \rangle$ dimer chain (Group I),
a surface with only the $\langle 110 \rangle$ dimer chain (Group II),
and a surface with both the $\langle 310 \rangle$ and 
the $\langle 110 \rangle$ dimer chains (Group III).
The relation between these groups and the combination of elements is summarized in Table \ref{tbl:group}.
The Sn/Ge(001) surface belongs to Group III as well as the Si/Si(001) surface and the Ge/Si(001) surface.
The Si/Si(001) and the Ge/Si(001) surfaces are one of the most important and fundamental surfaces
of semiconductor industry.
It is intriguing that the surface structure of the Sn/Ge(001) surface is the same as that of these surfaces.

\begin{table}
\begin{center}
\caption{Relation between the kind of dimer chain and group-IV elements on Si and Ge(001) surfaces. 
There are three kinds of surfaces:
a surface with only the $\langle 310 \rangle$ dimer chain (Group I),
a surface with only the $\langle 110 \rangle$ dimer chain (Group II), 
and a surface with both the $\langle 310 \rangle$
and the $\langle 110 \rangle$ dimer chains (Group III).
The data were taken from references
\cite{galea2000,wulfhekel1997,qin1997,wingerden1997,veuillen1996,glueckstein1998,yang1995}.}

\begin{tabular}{cll}
    \hline
    \hline
    Group & Kind of dimer chain & Surface \\
    \hline
    I & $\langle 310 \rangle$ dimer chain & Ge/Ge(001), Si/Ge(001)\\
    II & $\langle 110 \rangle$ dimer chain & Pb/Si(001), Sn/Si(001), Pb/Ge(001)\\
    III & $\langle 310 \rangle$ and $\langle 110 \rangle$ dimer chains & Ge/Si(001), Si/Si(001), Sn/Ge(001)$^{*}$\\
    \hline
    \multicolumn{3}{l}{$^{*}$Present work}\\
\end{tabular}
\label{tbl:group}
\end{center}
\end{table} 
\par

\begin{figure}[htbp]
    \begin{center}
        \includegraphics[keepaspectratio=true,scale=0.6]{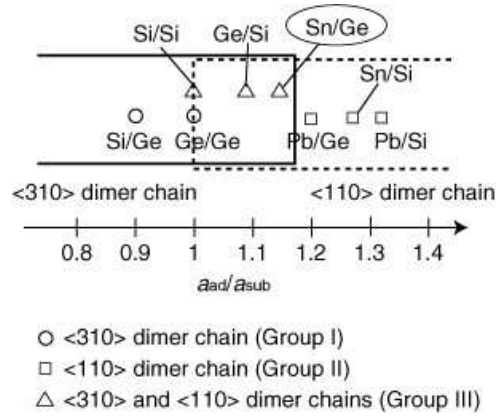}
    \end{center}
    \caption{Relation between the lattice constant difference
    and the kind of dimer chain. 
    The horizontal axis indicates the ratio of the covalent radii $a_{ad}/a_{sub}$,
    where $a_{ad}$ and $a_{sub}$ are covalent radii of the adatom and the substrate atom, respectively.
    The surfaces that belong to group I, II, and III are indicated by
    circles, rectangles, and triangles, respectively. 
    For the covalent radii of Si, Ge, Sn, and Pb atoms, 
    we used $1.11 \textrm{\AA}$, $1.22 \textrm{\AA}$, $1.41 \textrm{\AA}$, 
    and $1.47 \textrm{\AA}$, respectively \cite{winter2006}.
    When the covalent radius of the adatom is larger than that of the substrate atom, 
    the $\langle 110 \rangle$ dimer chain is formed. 
    In the opposite case, the $\langle 310 \rangle$ dimer chain is formed.
    On the other hand, when the covalent radius of the adatom is
    almost the same as that of the substrate atom,
    the $\langle 110 \rangle$ and the $\langle 310 \rangle$ dimer chains coexist.}
    \label{fig:strain}
\end{figure}

We can interpret the results in Table \ref{tbl:group}
in terms of lattice strain induced by the mismatch between the substrate and the ad-dimers.
Figure \ref{fig:strain} shows how the kind of dimer chain depends on the ratio of covalent radii.
We plotted results for the Group I, II, and III
by circles, rectangles, and triangles, respectively, with respect to the ratio.
It is reasonable to consider the covalent radius as the index of the strain
because the ad-dimer and the substrate atoms form a covalent bond
as discussed in Section \ref{sec:formation}.
We can see that the $\langle 110 \rangle$ ($\langle 310 \rangle$) dimer chain tends to be formed when the adatom is larger (smaller) than the substrate atom. 
On the other hand, when the size of adatom is almost the same as that of the substrate atom,
both of the dimer chains are formed.
These suggest that the lattice strain induced by the adatom determines the kind of dimer chain.

\par

In the previous section, we have discussed that the $\langle 110 \rangle$ dimer chain
is formed through the diffusion process of Sn adatoms,
which overcome the diffusion barrier to the edge site of the dimer chain.
In general, the bulk cohesive energy decreases with increasing the atomic radius,
and therefore the bond strength between the substrate atom and the adatom becomes
weaker with increasing their radius.
On the other hand, the diffusion barrier height relates to the bond strength
between the adatom and the substrate atom: stronger bond results in 
higher diffusion barrier. 
It has been reported by \textit{ab-initio} calculations that
diffusion barrier of a Si adatom on the clean Si(001) surfaces
is higher than that of a Ge or Si adatom on the clean Ge(001) surfaces
in accordance with the above discussion \cite{huang2004}.
Thus, the diffusion barrier from the site 2 to 1 in Fig. \ref{fig:qin} seems to decrease (increase)
when the covalent radius of the adatom is larger (smaller) than that of the substrate atom.
\par

This accounts for the experimental results that only the $\langle 110 \rangle$ dimer chains
tend to be formed for the Pb adatom.
However, this interpretation for diffusion barrier in terms of the bond strength is not perfect.
Only the $\langle 310 \rangle$ dimer chains are formed for the Ge/Ge(001) (Si/Ge(001)) surfaces,
and both $\langle 310 \rangle$ and $\langle 110 \rangle$ dimer chains
are formed for the Si/Si(001) surfaces \cite{galea2000,wulfhekel1997,wingerden1997}.
On the other hand, the interpretation in terms of lattice strain can give suitable explanation
for all the surfaces without exception as shown in Fig. \ref{fig:strain}. 
The kinetic barrier height seems to be governed by the lattice strain rather than by the bond strength.
In the present study,
how the strain, instead of the bond strength, affects the kinetic barrier height remains unanswered.
In order to gain the insights on the mechanism of the strain-mediated formation process of the dimer chains,
further research including theoretical approach will be needed.

\section{Summary}\label{sec:summary}
We have investigated the initial growth of the Sn atoms on the Ge(001) surface by means of STM.
For Sn deposition onto the substrate at RT, similarly to Si/Si(001) and Ge/Si(001) surfaces,
the Sn atoms form two kinds of dimer chains with different alignment:
the $\langle 310 \rangle$ and the $\langle 110 \rangle$ dimer chains.
The epitaxial structures composed of three ad-dimers (the triplet dimers) are also formed. 
The Sn ad-dimer in the $\langle 310 \rangle$ dimer chain is intrinsically symmetric or slightly buckled,
aligning in the $\langle 310 \rangle$ direction.
On the other hand, that in the $\langle 110 \rangle$ dimer chain is intrinsically asymmetric,
aligning in the $\langle 110 \rangle$ direction.
The ad-dimers in both of the dimer chains are located in the trough between the Ge dimer rows,
and have dimer axes parallel to the substrate Ge dimer's axis.
\par

For Sn deposition at 80 K, the population of the $\langle 310 \rangle$ dimer chains 
relative to that of the $\langle 110 \rangle$ dimer chains increases.
This is attributed to the existence of the diffusion barrier for the Sn adatoms
for the formation of the $\langle 110 \rangle$ dimer chain.
This is consistent with the model proposed by Qin \textit{et al}. for Si/Si(001) and Ge/Si(001) surfaces.
We propose that the diffusion barrier height for the Sn adatoms on the substrate
depends on surface strain induced by the Sn ad-dimer. 
For the $\langle 110 \rangle$ dimer chain,
the ad-dimer buckling lowers the diffusion barrier height.
The two kinds of dimer chains appearing on the Ge(001) and Si(001) surfaces
with adatoms of the group-IV elements are systematically interpreted in terms of stain 
induced by the ad-dimers on the substrate.

\section*{Acknowledgement}
The authors are grateful to Yasumasa Takagi, Masamichi Yamada, and Shunsuke Doi for helpful discussion.

\end{document}